\documentclass[aps,prb,twocolumn,showpacs,superscriptaddress,10pt,a4paper]{revtex4-1}
\usepackage{graphicx}
\usepackage{amsmath}
\usepackage{amssymb}
\usepackage{verbatim}
\usepackage{color}

\usepackage{amsthm}

 \theoremstyle{definition}

 \theoremstyle{remark}

\begin{document}

\title{Physical solutions of the Kitaev honeycomb model}

\author{Fabio L. Pedrocchi}
\affiliation{Department of Physics, University of Basel, Klingelbergstrasse 82, CH-4056 Basel, Switzerland}

\author{Stefano Chesi}
\affiliation{Department of Physics, University of Basel, Klingelbergstrasse 82, CH-4056 Basel, Switzerland}
\affiliation{Department of Physics, McGill University, Montreal, Quebec, Canada H3A 2T8}

\author{Daniel Loss}
\affiliation{Department of Physics, University of Basel, Klingelbergstrasse 82, CH-4056 Basel, Switzerland}

\begin{abstract}
We investigate the exact solution of the honeycomb model proposed by Kitaev and derive an explicit formula for the projector onto the physical subspace. The physical states are simply characterized by the parity of the total occupation of the fermionic eigenmodes. We consider a general lattice on a torus and show that the physical fermion parity depends in a nontrivial way on the vortex configuration and the choice of boundary conditions. In the vortex-free case with a constant gauge field we are able to obtain an analytical expression of the parity. For a general configuration of the gauge field the parity can be easily evaluated numerically, which allows the exact diagonalization of large spin models. We consider physically relevant quantities, as in particular the vortex energies, and show that their true value and associated states can be substantially different from the one calculated in the unprojected space, even in the thermodynamic limit. 

\end{abstract}


\maketitle
\section{INTRODUCTION}\label{sec:int}

The Kitaev honeycomb model, with several variations, has attracted a lot of attention over the last years. \cite{OpticalLattices, KitaevHoney, ShankarPRL2007,Vidal2008,  Pachos2008,PachosPRL2008,FiniteSizeEffect2009,KitaevMott,ChaloupkaPRL2010, Nori,KitaevPRL2011,MemoryPRB,ChesiPRA,KitaevToric, MandalPRB2009,Taylor2011,Trebst2011,Schmidt2010,YaoPRL2007,YaoPRL2009,YaoPRL2010,Willans2011,Nussinov2008,KellsPRB2009,Lahtinen2001New} Many different interesting aspects of it have been studied in detail in the original work of Kitaev. \cite{KitaevHoney} There, an exact method of solution of the model based on the mapping to Majorana fermion operators is discussed. Although alternative mappings and approximation techniques exist,\cite{Vidal2008,Nussinov2008,KellsPRB2009} Kitaev's method is widely applied, being ideally suited to this class of spin models. Further, the presence of an abelian and a non-abelian phase (in the presence of an external magnetic perturbation) was demonstrated. \cite{KitaevHoney} 
The Kitaev honeycomb model has a wide spectrum of physical applications, ranging from the description of strongly correlated materials \cite{KitaevMott} to the analytical study of critical quantum spin liquids. \cite{KitaevPRL2011} It is also of central importance in the context of quantum information theory since its gapped phase provides a perturbative realization of the toric code. \cite{KitaevToric} Extensions of the honeycomb model have been lately proposed as promising candidates for the realization of a topological quantum memory. \cite{ChesiPRA, MemoryPRB} Although very challenging, its physical realization has become closer to reality thanks to recent proposals. \cite{Nori, OpticalLattices}

In this  paper  we examine the projection to the physical subspace of the exact mapping to Majorana fermions proposed by Kitaev. \cite{KitaevHoney} As briefly discussed in Refs. \onlinecite{YaoPRL2007,YaoPRL2009,YaoPRL2010}, unprojected and projected models have different physical properties, especially the parity of fermions.   We derive here for the first time an explicit and immediately applicable representation of the projector in terms of the parity of physical fermions. As it turns out, the physical fermion parity depends in a nontrivial way on the configuration of vortices and on the lattice topology. 
 Applying the projection to specific cases, we  find large differences between projected and unprojected physical quantities (e.g. ground state and vortex energies, or spin-spin correlation functions). Such discrepancies exist both in the gapped and gapless phase and can also survive the thermodynamic limit. Our analysis is consequently essential for the exact numerical study via Kitaev's exact mapping of large spin systems, especially if one wants to go beyond the small system sizes of about $20-100$ spins currently accessible to various numerical approaches such as direct diagonalization \cite{Pachos2008,FiniteSizeEffect2009,ChaloupkaPRL2010} or density matrix renormalization group (DMRG). \cite{Trebst2011} These numerical approaches become necessary if exact fermionization techniques are not applicable.

The paper is organized as follows. In Sec. \ref{sec:model} we briefly review the honeycomb model and the exact mapping to Majorana fermions introduced by Kitaev.\cite{KitaevHoney} In Sec. \ref{sec:reduced} we compute the parity of physical fermions with periodic boundary conditions and a generic vortex configuration, which represents  our main result. Section \ref{sec:numerical} contains some applications to specific cases and Sec. \ref{sec:conclusion} our final remarks.

\section{Model and exact mapping}\label{sec:model}

\begin{figure}
	\centering
		\includegraphics[width=0.47 \textwidth]{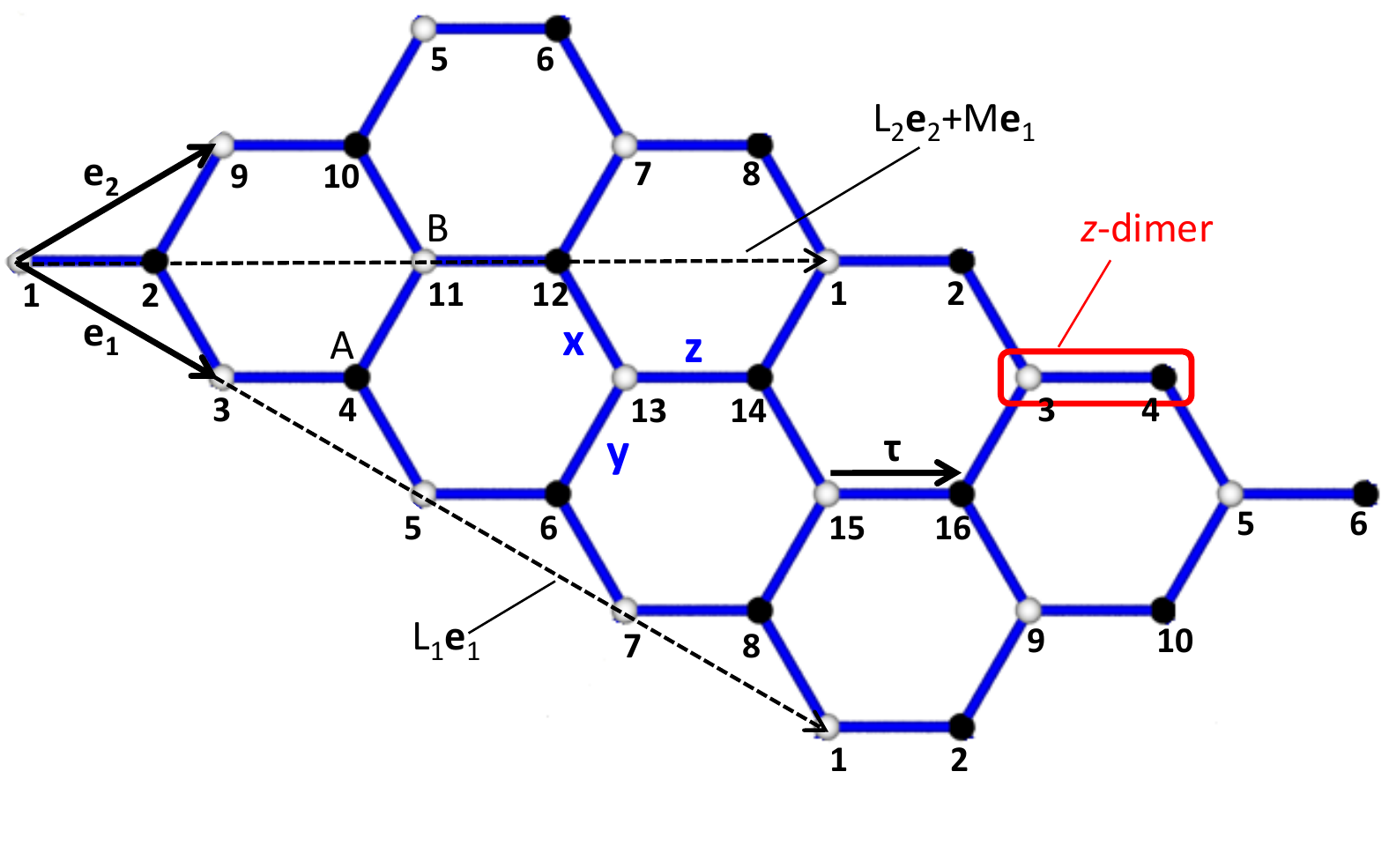}
	\caption{Honeycomb lattice, with basis vectors ${\bf e}_{1,2}$. The directions of $x$, $y$, $z$-links are indicated, as well as the vector $\boldsymbol{\tau}$ joining \color{black} B and A \color{black} sublattices (white and black dots, respectively). The most general torus on the lattice can be specified by $L_1 {\bf e}_1$ and $L_2 {\bf e}_2+ M{\bf e}_1$ (here $L_{1}=4$, $L_{2}=M=2$). The numbers $1,2, \ldots 16$ label the sites as described in the main text, see Eq.~(\ref{eq:labeling}).}
	\label{fig:HoneycombTorus}
\end{figure}

The Kitaev honeycomb model is a quantum compass model \cite{Kugel} defined on an hexagonal lattice $\Lambda$ as follows
\begin{equation}\label{eq:Hamiltonian}
H=\sum_{\langle i,j \rangle} J_{\alpha_{ij}} \sigma^{\alpha_{ij}}_i \sigma^{\alpha_{ij}}_j,
\end{equation}
where $\boldsymbol{\sigma}_i$ are the Pauli spin operators at site $i \in \Lambda$ ($i=1, \ldots, 2N$). In Eq.~(\ref{eq:Hamiltonian}), the sum runs over all the pairs of nearest-neighbor sites and the directions of the Ising interactions are determined by the orientations of the corresponding links ($\alpha_{ij}=x,y,z$ for $x$-, $y$-, $z$-links respectively, see Fig.~\ref{fig:HoneycombTorus}).

To solve this spin model in an extended Hilbert space $\widetilde{\mathcal{L}}$, one can associate at each site $i$ four Majorana modes $c_{i},b_{i}^{x},b_{i}^{y},b_{i}^{z}$. \cite{KitaevHoney} By defining $\widetilde{\sigma}_{i}^{\alpha}=ib_{i}^{\alpha}c_{i}$, the original Hamiltonian in Eq.~(\ref{eq:Hamiltonian}) is mapped to 
\begin{equation}\label{eq:HamiltonianExtended}
\widetilde{H}=i \sum_{\langle i,j \rangle}\widehat{A}_{ij}c_{i}c_{j},
\end{equation}
where for nearest-neighbor sites $\widehat{A}_{ij}= J_{\alpha_{ij}} \widehat{u}_{ij} $ and
\begin{equation}
\widehat{u}_{ij} =  i b_{i}^{\alpha_{ij}}b_{j}^{\alpha_{ij}}=-\widehat{u}_{ji}.
\end{equation}
These operators satisfy $\widehat{u}_{ij}^2=1$. Furthermore, they all commute with each other and also with $\widetilde{H}$. Therefore, the extended Hilbert space splits into $\widetilde{\mathcal{L}}=\oplus_{u}\widetilde{\mathcal{L}}_{u}$, where $u$ represents a configuration of $u_{ij }=\pm 1$.  Notice that $u_{ij}=-u_{ji}$. So, whenever we specify the values of $u_{ij}$, we assume conventionally that $i$ is in the A sublattice (see Fig.~\ref{fig:HoneycombTorus}).  In each subspace $\widetilde{\mathcal{L}}_{u}$,  the operator matrix $\widehat{A}_{ij}$ are replaced by numbers $A^{u}_{i,j}$ and Eq.~(\ref{eq:HamiltonianExtended}) thus describes non-interacting Majorana fermions. 

The eigenmodes can be easily obtained with a canonical transformation $Q^u$ to new Majorana operators
\begin{equation}\label{eq:Q}
(b_{1}^{'},b_{1}^{''},...,b_{N}^{'},b_{N}^{''})=(c_{1},...,c_{2 N} ) Q^u,
\end{equation}
which for the specific configuration $u$ brings $\widetilde{H}$ to the form $\widetilde{H}_u=\frac{i}{2}\sum_{m}\epsilon_{m}b_{m}^{'}b_{m}^{''}$, where $\epsilon_{m}$ are the positive eigenvalues of $2iA^u$. By introducing the fermion operators $a_{m}=1/2(b_{m}^{'}+ib_{m}^{''})$  and $n_{m}=a_{m}^\dagger a_{m}$ we obtain
\begin{equation}\label{eq:Hamiltonian4}
\widetilde{H}_u=\sum_{m}\epsilon_{m}\left( n_m-1/2 \right), 
\end{equation}
with ground state energy $E_{0}=-1/2\sum_{m}\epsilon_{m}$. The orthogonal matrix  $Q^u$ will have a crucial role in the following to obtain the projection operator.
\section{Physical fermion parity}\label{sec:reduced}

The key advantage of Kitaev's solution is to reduce the problem of finding the eigenvalues of a $2^{2N}\times 2^{2N}$ matrix to the diagonalization of the $2N\times 2N$ matrices $A^u$. However, the final spectrum and eigenstates are in the extended Hilbert space $\widetilde{\mathcal{L}}$, and a projection $\mathcal P$ to the physical subspace is necessary. \cite{KitaevHoney} The physical states satisfy $D_{i}\vert\Psi\rangle=\vert\Psi\rangle$ for all the gauge operators $D_{i}=b_{i}^{x}b_{i}^{y}b_{i}^{z}c_{i}$ and the explicit form of $\mathcal{P}$ is \cite{KitaevHoney}
\begin{equation}\label{P_def}
\mathcal{P}=\prod_{i=1}^{2N}\left(\frac{1+D_{i}}{2}\right)=\frac{1}{2^{2N}}\sum_{\{i\}}\prod_{i \in\{i\}}D_i,
\end{equation}
where the summation runs over all possible subsets of indices $\{i\}\subseteq \Lambda$. Within the physical subspace the $\widetilde{\sigma}^{x,y,z}$ operators satisfy the usual algebra of Pauli matrices and therefore $H$ and $\widetilde{H}$ are equivalent. 

To establish a more explicit formula for $\mathcal{P}$ one can note that, in the summation appearing in Eq.~(\ref{P_def}), the two terms corresponding to a subset $\{ i \}$ and its complementary set $\Lambda\setminus \{i\}$ simply differ by a factor $\prod_{i=1}^{2N} D_i$. Therefore, $\mathcal{P}$ factorizes as follows \cite{YaoPRL2009,YaoPRL2010}:
\begin{equation}\label{P_factorization}
\mathcal{P}=\left(\frac{1}{2^{2N-1}}{\sum_{\{ i \}}}^\prime\prod_{i\in\{ i\}}D_i\right)\cdot \left(\frac{1+\prod_{i=1}^{2N}D_i}{2}\right)= \mathcal{S}\cdot \mathcal{P}_0,
\end{equation}
where the prime indicates that the summation in $\mathcal{S}$ (in the first parenthesis) is restricted to half of all possible subset of indices: if $\{ i \}$ is included, then $\Lambda \setminus  \{i \}$ is not. 

We then consider $\prod_{i=1}^{2N}D_i$ in the projector $\mathcal{P}_0$ [the second parenthesis of Eq.~(\ref{P_factorization})]. From the definition of $D_i$, it clearly consists of a product of all the $c_i$ and $b_i^{x,y,z}$ operators. By applying the anticommutation rules we can pair corresponding $b_i^{x,y,z}$ operators, and express them in terms of the conserved quantities $u_{ij}$. To do this, it is necessary to know the topology of the lattice, from which the correct pairing is determined. We consider here a model defined on a torus with basis vectors $L_1 {\bf e}_1$ and $L_2 {\bf e}_2+M {\bf e}_1$, as illustrated in Fig.~\ref{fig:HoneycombTorus} for a special case ($L_{1}=4$, $L_{2}=M=2$)\color{black}. This represents the most general choice of periodic boundary conditions and $N=L_1 L_2$. It is also necessary to fix the correspondence between $i=1,...,2N$ and lattice sites. We fix the labeling as in Fig.~\ref{fig:HoneycombTorus}. By taking the origin on site $i=1$ (on the $B$ sublattice) the position ${\bf r}_i$ of the sites with odd values of $i$ is given by
\begin{equation}\label{eq:labeling}
{\bf r}_i=
\left(\frac{i-1}{2} \bmod L_1 \right){\bf e}_1 
+ \left(\frac{i-1}{2}\backslash L_1 \right){\bf e}_2,
\end{equation}
where $(a \backslash b)$ indicates the integer division and $(a \bmod b)$ the reminder. For even $i$, the position is ${\bf r}_{i}={\bf r}_{i-1}+\boldsymbol{\tau}$ (see Fig.~\ref{fig:HoneycombTorus}). \color{black} After pairing the $b_i^{x,y,z}$ operators into the $u_{ij}$\color{black}, the result is proportional  to the parity operator $\hat\pi_c=(-i)^N \prod_i c_i$. We then express this quantity in terms of the eigenmodes:\begin{equation}\label{c_product}
\hat{\pi}_{c}=\det(Q^{u})\,\hat{\pi},
\end{equation}
where $\hat{\pi}=\prod_{m=1}^{N}(1-2n_{m})$ is the parity of the eigenmodes $a_{m}$. A proof of Eq.~(\ref{c_product}) is provided in Appendix~\ref{app1}. \color{black} Finally, we find for $\mathcal{P}_0$, 
\begin{equation}\label{eq:P0}
2\mathcal{P}_{0}=1+(-1)^{\theta}\det(Q^u)\,\hat{\pi}\prod_{\langle i,j\rangle}u_{ij},
\end{equation}
where $\theta=L_{1}+L_{2}+ M(L_{1}-M)$  \color{black} and $\hat{\pi}$ has eigenvalues $+1(-1)$ if the total number of \emph{physical} fermions is even (odd). A complete derivation of Eq.~(\ref{eq:P0}) is given in Appendix~\ref{app2}. It is important to notice that, in applying Eq.~(\ref{eq:P0}), the labeling of the lattice described above should be used. For example, $\det(Q^u)$ depends on the choice of the labeling. \color{black}

We would like to point out now the differences between Eq.~(\ref{eq:P0}) and other discussions in the literature. \cite{YaoPRL2007,YaoPRL2009,YaoPRL2010,Willans2011} Firstly, the parity of physical fermions (the only relevant ones) $\hat{\pi}$ is the parity of the eigenmodes $a_{m}$ and not simply $\hat{\pi}_c$ which is of no calculational use. Eq.~(\ref{c_product}) shows now the precise relation between $\hat{\pi}$ and $\hat{\pi}_{c}$: for a certain configuration $u$, the two parities are different if $\det(Q^u)=-1$.  We note also that $\hat\pi_c$ is not a gauge invariant quantity, while the physical parity $\hat\pi$ obviously is, i.e., $[\hat{\pi},D_{i}]=0$. Therefore Eq.~(\ref{c_product}) allows one  to form the gauge invariant quantity $\det(Q^u)\prod_{\langle i,j\rangle}u_{ij}$. A second feature revealed by our analysis is that $\hat\pi$ (and $\hat\pi_c$ as well) depends in a nontrivial way on the boundary conditions through the factor $(-1)^{\theta}$, which does not appear in Refs.~\onlinecite{YaoPRL2010,Willans2011}.

Being directly applicable to the eigenstates $|\Psi\rangle_u \in \widetilde{\mathcal{L}}_u$, identified by their occupation numbers $(1-2n_m)=\pm 1$, Eq.~(\ref{eq:P0}) is extremely convenient: it immediately shows whether $\mathcal{P}_0$ gives 0 or 1 on the eigenstate $|\Psi\rangle_u$. In the former case, the state is clearly unphysical. In the second case, $\mathcal{P}|\Psi\rangle_u = \mathcal{S}|\Psi\rangle_u \neq 0$ since, as seen in Eq.~(\ref{P_factorization}), the $2^{2N-1}$ terms of $\mathcal{S}$ all correspond to different configurations of $u_{ij}$. In conclusion, Eq.~(\ref{eq:P0}) is sufficient to determine if $|\Psi\rangle_u$ has some overlap with the physical subspace or lies completely outside of it, and makes clear that the crucial quantity is the parity of \emph{physical} fermions $\hat{\pi}$: physical states have either even or odd occupation of the eigenmodes $a_{m}$ depending on \emph{both} the configuration $u$ and the choice of boundary condition.  

\section{Examples of projected states and energies}\label{sec:numerical}

We discuss in this section a few examples illustrating the difference between physical and unphysical results. These examples should make clear that it is not sufficient to calculate the properties of the system in the extended space, but the projection must be carefully applied to take advantage of all the power of Kitaev's exact mapping. While we focus here on the ground state and vortex excitation energy, we expect that similar discrepancies exist for other physical quantities. 

By applying  Eq.~(\ref{eq:P0}), the only factor which is not immediately found is $\det(Q^u)$ and we show in the following how it can be explicitly evaluated when $u_{ij}=1$. The final result, Eq.~(\ref{eq:determinant}), nicely complements Eq.~(\ref{eq:P0}) for this vortex-free sector. 

For an arbitrary configuration of the $u_{ij}$, the Fourier transformation cannot be used to calculate analytic results. However, $\det(Q^{u})$ can be determined numerically with negligible computational effort. This allows us to obtain the exact numerical solution of the spin Hamiltonian at very large values $N$ and to explore the effect of the projection when approaching the thermodynamic limit. 

\subsection{Vortex-free sector}

\begin{figure}
	\centering
		\includegraphics[width=0.4 \textwidth]{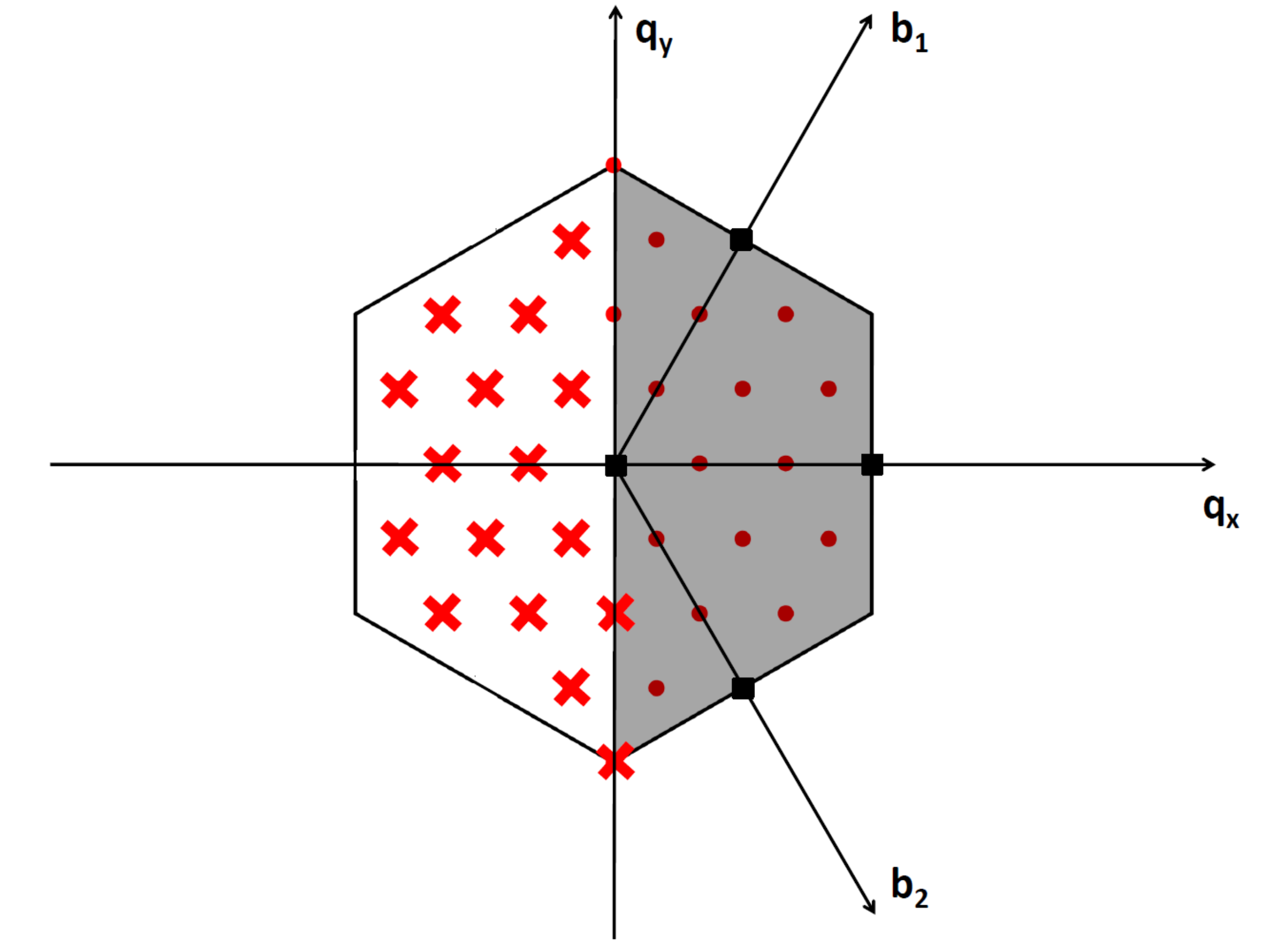}
	\caption{An illustration of the partitioning into $\Omega,\Omega_\pm$ described in the main text. The vectors ${\bf b}_{1,2}$ define the reciprocal lattice and the hexagon is the first Brillouin zone. The four square points are wavevectors in $\Omega$ while the dots and crosses are in $\Omega_+$ and $\Omega_-$, respectively. In this example $L_1=L_2=6$ and $M=0$.}
	\label{fig:Omegas}
\end{figure}

In the vortex-free sector with $u_{ij}=1$ we can proceed as in Ref.~\onlinecite{KitaevHoney} by making use of the Fourier transform on the lattice
\begin{eqnarray}\label{FTransf}
&& a_{{\bf q}B}=\frac{1}{\sqrt{2N}}\sum_{i~\rm odd}e^{-i{\bf q}\cdot{\bf r}_{i}}c_{i}, \nonumber \\
&& a_{{\bf q}A}=\frac{1}{\sqrt{2N}}\sum_{i~\rm odd}e^{-i{\bf q}\cdot{\bf r}_{i}}c_{i+1}, 
\end{eqnarray}
where the positions ${\bf r}_i$ of the $B$ lattice sites are given in Eq.~(\ref{eq:labeling}). \color{black}  
We now consider a partition of the $N$ possible values of ${\bf q}$ (fixed by the periodicity of the lattice) in three sets: $\Omega$ and $\Omega_\pm$. We say that ${\bf q}\in \Omega$ if $\pm{\bf q}$ are the same (up to reciprocal lattice vectors). $\Omega$ contains at most four wave vectors, depending on $L_{1,2}$ and $M$,  and always contains ${\bf 0}$. The remaining wavevectors can be partitioned in a way that $\pm{\bf q}$ always belong to two distinct sets $\Omega_\pm$. An example of such partitioning is illustrated in Fig.~\ref{fig:Omegas} for the special case $L_1=L_2=6$ and $M=0$. We define new Majorana modes as $\gamma_{\bf q}^\lambda=\sqrt{2}a_{{\bf q}\lambda}$ for ${\bf q}\in\Omega$ and
\begin{equation} 
\gamma_{{\bf q},1}^\lambda=a_{{\bf q}\lambda}+a_{{\bf -q}\lambda}, \quad
\gamma_{{\bf q},2}^\lambda=i (a_{{\bf q}\lambda}-a_{{\bf -q}\lambda}),
\end{equation}
for ${\bf q} \in \Omega_+$, where $\lambda=A,B$ refers to the sublattice (see Fig.~\ref{fig:HoneycombTorus}). \color{black} This canonical transformation of the $c_{i}$ can be constructed in two steps. First we rearrange the $(c_1,c_2,\ldots, c_{2N})$ into
\begin{equation}\label{eq:reordering}
(c_2, c_4, \ldots , c_{2N}, c_1, c_3 , \ldots ,c_{2N-1}),
\end{equation}
a transformation which has determinant $(-1)^{N(N+1)/2}$. Notice that, in Eq.~(\ref{eq:reordering}), the modes are partitioned between $A$ (first half) and $B$ (second half). Furthermore, for each sublattice the same order of z-dimers appears, since $c_i$ and $c_{i+1}$ (with odd $i$) belong to the same z-dimer (see Fig.~\ref{fig:HoneycombTorus}). Because of this structure, the second transformation [from Eq.~(\ref{eq:reordering}) to the $\gamma^\lambda_{\bf q},\gamma^\lambda_{\bf q,\alpha}$ modes]
\color{black}
has two identical blocks labeled by $\lambda = {\rm A, B}$ and the determinant is simply 1.

The Hamiltonian, rewritten in terms of the new Majorana modes, is diagonal in ${\bf q}$ and its coefficients are given by $f({\bf q})=2(J_{x}e^{i{\bf q}\cdot{\bf e}_{1}}+J_{y}e^{i{\bf q}\cdot{\bf e}_{2}}+J_{z})$ and its complex conjugate. A further diagonalization with respect to the index $\alpha$ of $\gamma_{{\bf q},\alpha}^{\lambda}$ is achieved with the rotation of the $B$ operators: \color{black}
\begin{equation}
\begin{pmatrix}
\widetilde{\gamma}_{{\bf q},1}^{B}\\
\widetilde{\gamma}_{{\bf q},2}^{B}
\end{pmatrix}=\begin{pmatrix}
              \cos(\phi_{{\bf q}}) & \sin(\phi_{{\bf q}})\\
              -\sin(\phi_{{\bf q}}) & \cos(\phi_{{\bf q}})\end{pmatrix}\begin{pmatrix}\gamma_{{\bf q},1}^{B}\\
              \gamma_{{\bf q},2}^{B}\end{pmatrix}
\end{equation}
where $\phi_{{\bf q}}$ is the phase of $f({\bf q})$, i.e., $f({\bf q})=\vert f({\bf q})\vert e^{i\phi_{{\bf q}}}$. This transformation has again determinant 1 and brings $\widetilde{H}_{u}$ to
\begin{equation}\label{eq:vsdnj}
\widetilde{H}_{u}=\frac{i}{2}\left(\sum\limits_{{\bf q}\in\Omega_+}\vert f({\bf q})\vert \sum_{\alpha=1,2}\gamma_{{\bf q},\alpha}^{A}\widetilde{\gamma}_{{\bf q},\alpha}^{B}+\sum_{{\bf q}\in\Omega} f({\bf q})\gamma_{{\bf q}}^A\gamma_{{\bf q}}^B \right).
\end{equation}
Finally, as discussed below Eq.~(\ref{eq:Q}), $Q^u$ brings the Hamiltonian to the form $\widetilde{H}_{u}=\frac{i}{2}\sum_{{\bf q}}\epsilon({\bf q})b_{{\bf q}}^{'}b_{{\bf q}}^{''}$ with $\epsilon({\bf q}) \geq 0$. This can be achieved in Eq.~(\ref{eq:vsdnj}) by relabeling the Majorana operators. If ${\bf q}\in\Omega_+$:
\begin{eqnarray}
b_{{\bf q}}^{'}=\gamma_{{\bf q},1}^{A}&, &b_{{\bf q}}^{''}=\widetilde{\gamma}_{{\bf q},1}^{B},\\
b_{-{\bf q}}^{'}=\gamma_{{\bf q},2}^{A}&, &b_{-{\bf q}}^{''}=\widetilde{\gamma}_{{\bf q},2}^{B}.
\end{eqnarray}
If ${\bf q}\in \Omega$ and $f({\bf q})\geq 0(< 0)$: $b_{\bf q}^\prime = \gamma_{\bf q}^{A(B)}$, $b_{\bf q}^{''} = \gamma_{\bf q}^{B(A)}$. The determinant of this last transformation is $(-1)^{\chi+N(N-1)/2} $, where $\chi$ is the number of reciprocal lattice vectors ${\bf q}\in\Omega$ such that $f({\bf q})<0$.  By combining this factor with the one from Eq.~(\ref{eq:reordering}) we obtain
\begin{equation}\label{eq:determinant}
\det{(Q^u)} = (-1)^{\chi+N^2} \quad {\rm for}\,\,\,\,\,~u_{ij}=1.
\end{equation}
\color{black}
Notice that $\chi$ depends in a non trivial way on the boundary conditions $L_{1,2},M$, and the couplings $J_{x,y,z}$. Nevertheless, for a given choice of the model, it can be easily computed.

\begin{figure}
	\centering
		\includegraphics[width=0.45\textwidth]{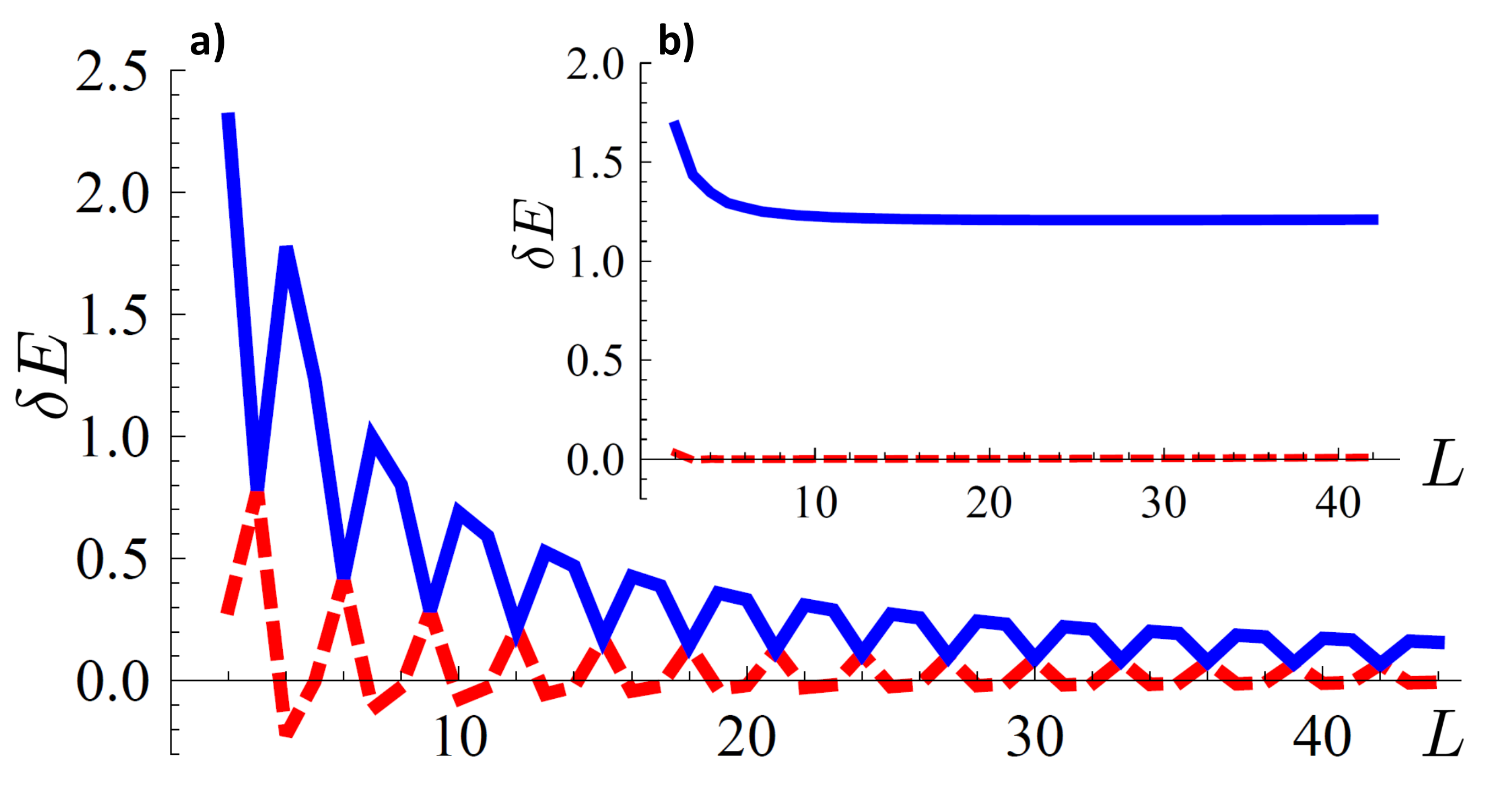}
	\caption{Physical (solid line) and unphysical (dashed line) finite size corrections to the thermodynamic energy of a vortex-free configuration $u_{ij}=1$ for $L_1=L_2=L$. The main plot $a)$ refers to the gapless phase with $M=0$, $J_x=J_y=J_z=1$, and $\varepsilon_{0}\simeq-1.5746$. \cite{KitaevHoney} The inset $b)$ corresponds to the gapped phase with $M=1$, $J_{x}=J_{y}=0.2$, $J_{z}=1$, and $\varepsilon_{0}\simeq -1.0202$.}
	\label{fig:KitaevGraphWithProjPaper}
\end{figure}

Following Ref.~\onlinecite{KitaevHoney} we examine now the finite size correction $\delta E(N) = E_0(N) - \varepsilon_0 N$ to the ground-state energy $E_0(N)$ with $u_{ij}=1$, where 
\begin{equation}
\varepsilon_{0} = \lim_{N \to \infty} E_0(N)/N
\end{equation}
is the energy per unit cell in the thermodynamic limit. We consider in Fig.~\ref{fig:KitaevGraphWithProjPaper} a square lattice ($L_1=L_2=L$) with $u_{ij}=1$ and two different choices of boundary conditions and couplings, and plot $\delta E(N)$ as a function of $L$ ($N=L^2$ in this case). The original result calculated in Ref.~\onlinecite{KitaevHoney} is reproduced in the main panel of Fig.~\ref{fig:KitaevGraphWithProjPaper} (dashed lines) and evidently refers to the unphysical energy, which always underestimates the correct result. The physical and unphysical energies are always distinct, unless $\epsilon_m =0$ for some fermion mode. Being in the gapless phase, the difference approaches zero at large system size as $1/L$. The inset represents an example in the gapped phase: remarkably, since there is always one fermion in the physical  ground state, the difference between projected (solid line) and unprojected (dashed line) results survive the thermodynamic limit where the true energy correction does not vanish.  That the state with zero fermions is never physical for any $L$ is immediate from our analytic result Eq.~(\ref{eq:determinant}) since $\chi=0$ in the gapped phase [$f({\bf q})>0$].
On the other hand, projected and unprojected \emph{states} are different (have different parity) in the thermodynamic limit in both the gapless and gapped phases. Therefore our projection protocol is necessary to determine the physical quantities of the model in  both the gapless and gapped  phase even in the thermodynamic limit. 

\subsection{Energy of two adjacent vortices}

\begin{figure}
	\centering
		\includegraphics[width=0.45\textwidth]{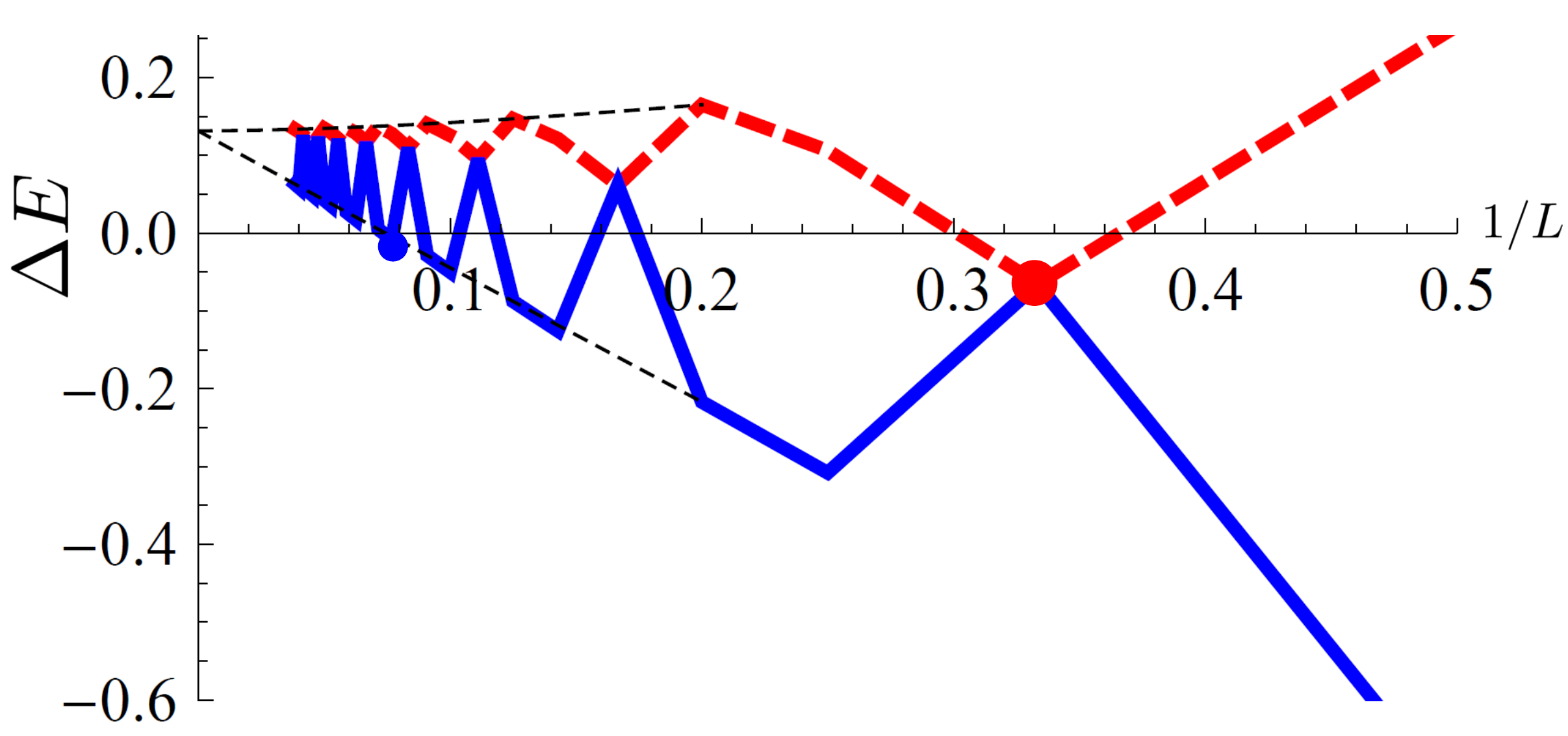}
	\caption{Physical (thick solid line) and unphysical (thick dashed line) excitation energy of two adjacent vortices. This plot refers to the gapless phase with $J_x=J_y=J_z=1$, $L_1=2L$, $L_2=M=L$, and a single $u_{ij}=-1$ with $ij$ being a $z$ link. Thin dashed lines extrapolate to the thermodynamic limit. The dots at $L=3$ (unphysical) and $L=13$ (physical) represent the largest system sizes with negative excitation energy.}
	\label{fig:Extrapolation_paper_1}
\end{figure}

\begin{figure}
	\centering
		\includegraphics[width=0.45\textwidth]{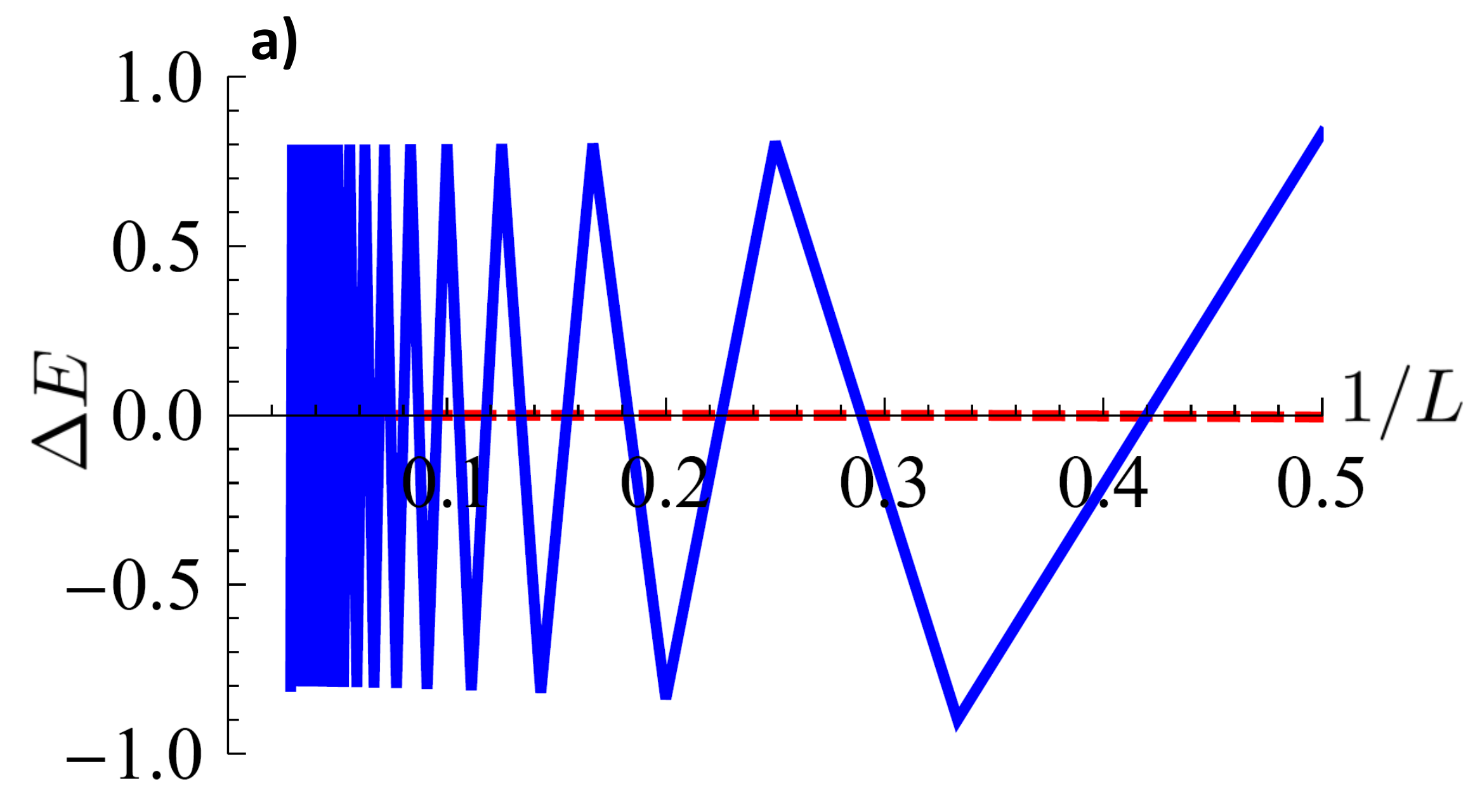}
		\includegraphics[width=0.45\textwidth]{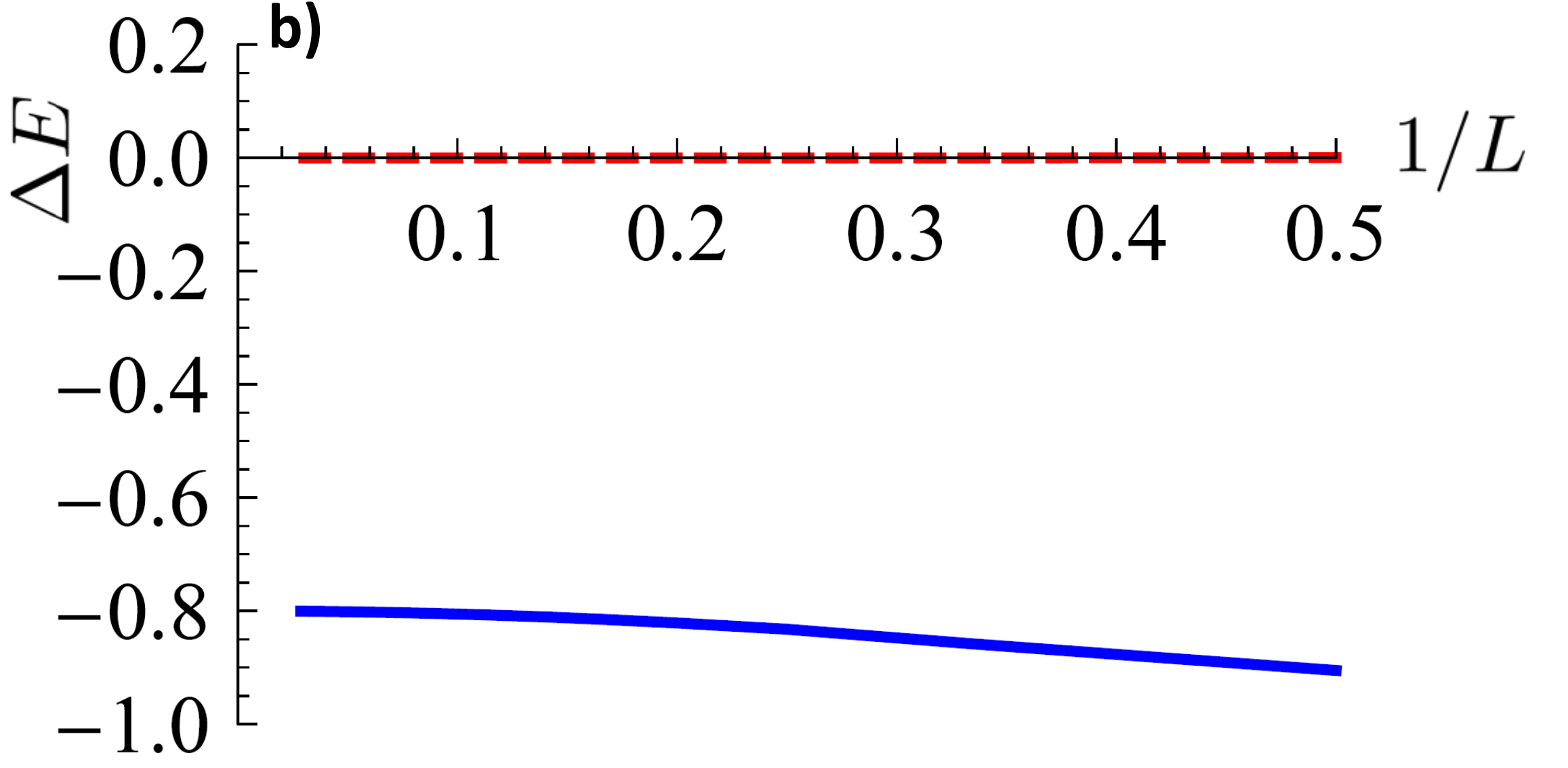}
	\caption{Physical (solid line) and unphysical (dashed line) excitation energy of two adjacent vortices in the gapped phase with $J_{x}=J_{y}=0.1$, $J_{z}=1$, $L_{1}=L_{2}=L$, and a single $u_{ij}=-1$ with $ij$ being a $y$ link.  Panel  $a)$ refers to $M=0$ while panel  $b)$ to $M=1$.}
	\label{fig:Extrapolation_paper_2}
\end{figure}

We consider next the energy to create vortices in the system. These are present on hexagonal cells for which the product of the six $u_{ij}$ is $-1$. As an interesting example we study configurations with two adjacent vortices, obtained by setting $u_{ij}=-1$ for a single link.  The ground state energy of such two-vortex configuration of the $u_{ij}$ is denoted as $E_2(N)$ while, as before, $E_0(N)$ is the ground energy with all $u_{ij}=1$. We define the excitation energy of a vortex as
\begin{equation}
\Delta E(N) =\frac12 [E_2(N)-E_0(N)],
\end{equation}
which is plotted in Figs.~\ref{fig:Extrapolation_paper_1} and \ref{fig:Extrapolation_paper_2} as function of $L$ for different choices of the parameters (given in the captions). In particular, Fig.~\ref{fig:Extrapolation_paper_1} refers to the gapless phase and Fig.~\ref{fig:Extrapolation_paper_2} to the gapped phase. Notice also that it is important in general to specify if $u_{ij}=-1$ refers to an $x$, $y$, or $z$ link.  

Since the vortex state is not translationally invariant, we obtain the energy spectrum  and $\det(Q_u)$  numerically. This can be done efficiently, since it only involves $2N\times 2N$ matrices. In Figs.~\ref{fig:Extrapolation_paper_1} and \ref{fig:Extrapolation_paper_2}  we show physical excitation energies up to $L=26$  and $L=35$, respectively  (corresponding to $4L^2=2704$ and $2L^2=2450$ \color{black} physical spins), obtained with a standard tabletop computer and high level language routines (\textsc{matlab}). We have checked that for the lowest possible system size the spectrum obtained with this method is in agreement with direct diagonalization of the physical model. However, direct diagonalization is limited to systems with only a few tens of spins, by making use of very intensive parallel computing numerical routines. \cite{Pachos2008, ChaloupkaPRL2010} Other numerical approaches such as DMRG\cite{Trebst2011}  allow to address larger spin systems ($2N\lesssim 100$)  but still much smaller than the ones accessible with the projection protocol presented here.


Similarly to Fig.~\ref{fig:KitaevGraphWithProjPaper}, physical and unphysical results are generally different. In Fig.~\ref{fig:Extrapolation_paper_1} (in the gapless phase) they become equal only if there is a zero energy mode or in the thermodynamic limit. On the other hand, finite size corrections are very important: $\Delta E(N)$ shows pronounced oscillations with an amplitude which is of the same order of magnitude of $\Delta E(\infty)$, for up to a few thousand spins. Remarkably, such oscillations result in negative excitation energies of the vortex pair up to $676$ spins (instead of $36$, in the extended space).

 In the gapped phase, the difference between projected and unprojected results can survive the thermodynamic limit. In Fig.~\ref{fig:Extrapolation_paper_2}a the oscillations in the physical solution (solid line) persist for $L\rightarrow\infty$ and their size is equal to half the gap of the fermions ($\sim J_{z}$), much larger than the excitation energy of the vortex pair in the unphysical space (dashed line). A different choice of boundary conditions can lead to a situation where $\Delta E$ is always large and negative ($\sim -J_z$)  and has a well defined thermodynamic limit, as illustrated in Fig.~\ref{fig:Extrapolation_paper_2}b.

The thermodynamic limit of the unphysical energies (dashed curves) of Fig.~\ref{fig:Extrapolation_paper_2} is well approximated by a high order perturbative expansion, which allows to derive an accurate effective Hamiltonian including vortex energies and interactions.\cite{KitaevHoney, Vidal2008,Schmidt2010, MemoryPRB} However such low-energy Hamiltonian does not contain explicitly the constraints on the allowed vortex configurations, and some care is necessary to establish which states are physical. For example, the vortex-free configuration of Fig.~\ref{fig:Extrapolation_paper_2}b does not belong to the low-energy subspace and the two-vortex state has always lower energy. It is worth pointing out here that the negative vortex energies encountered in these examples are not in contradiction with Lieb's theorem as originally formulated in Ref.~\onlinecite{Lieb} (see also Ref.~\onlinecite{Macris1996}).

\section{Conclusion}\label{sec:conclusion}

We have obtained here an explicit form of the projection operator which allows us to extract the physical properties of the honeycomb model for large lattices.  The parity of fermions in the physical sector is directly given through our Eq.~(\ref{eq:P0}) and depends in a nontrivial way on the vortex configuration and the periodic boundary conditions. By applying Eq.~(\ref{eq:P0}), we have examined the energies of vortex-free and two-vortex configurations and showed that significant deviation from the physical values can exist if the projection operator is not taken into account. Such differences between projected and unprojected quantities can persist up to large values of $N$ and they can survive the thermodynamic limit.

Applying the projection only requires to determine the parity of physical fermions. Therefore, it does not introduce any additional  complication  related to the symmetrization over all gauge transformations (\ref{P_def}), a procedure which never needs to be implemented in practice. As known, the energies of projected (i.e. symmetrized) and unprojected (i.e. unsymmetrized) states \color{black} are the same and
the spin correlation functions $S_{ij}^{\alpha\beta}(t)=\langle \sigma_i^\alpha(t)\sigma_j^\beta(0) \rangle$ (whose certain exact properties were discussed in Refs.~\onlinecite{ShankarPRL2007,Nussinov2008}) can be conveniently computed with unprojected eigenstates $|\Psi \rangle_u$.\cite{ShankarPRL2007} This is possible thanks to the fact that the spin operators $\widetilde \sigma_i^\alpha$ are gauge-invariant. However, one should make sure that only states with $\mathcal{P}|\psi \rangle_u \neq 0$ are included in $S_{ij}^{\alpha\beta}(t)$. Therefore deviations from the unprojected results exist for the $S_{ij}^{\alpha\beta}(t)$ as well. 

More generally it is obvious from our discussion that all dynamic and thermodynamic quantities derived from $H$ (for example, the partition function), depend on the physical spectrum and eigenstates and thus differ from those of the unprojected model $\widetilde H$. Therefore, we think that it would be interesting to apply our method to problems which have been studied without projection \cite{KitaevHoney,Taylor2011,Schmidt2010,Lahtinen2001New} and compare the differences. Our work is generally relevant to spin models to which Kitaev's solution applies, like the honeycomb model perturbed by a weak magnetic field \cite{KitaevHoney,Pachos2008,Trebst2011}, interacting with cavity modes \cite{MemoryPRB} or with different link distributions \cite{Schmidt2010}, and a three-dimensional extension of the honeycomb model recently proposed in Ref.~\onlinecite{MandalPRB2009}. The case of open boundary conditions can also be simply obtained by extending Eq.~(\ref{eq:Hamiltonian}) to site-dependent couplings $J_{\alpha_{ij}} \to J_{ij}$ (and $J_{ij}=0$ on the boundary).

\section{Acknowledgments}
We thank D. DiVincenzo for inspiring suggestions. We also acknowledge discussions with S. Gangadharaiah, V. Lahtinen, D. Rainis, B. R\"othlisberger, and L. Trifunovic. This work was supported by the Swiss NSF, NCCR Nanoscience, NCCR QSIT, DARPA QuEST, and the EU project SOLID. S.C. acknowledges support from CIFAR.

\appendix
\section{Derivation of Eq.~(\ref{c_product})\label{app1}}
In this appendix we present a detailed derivation of Eq.~(\ref{c_product}) which gives the relation between the parity $\hat{\pi}_{c}$ and the parity of the physical fermions $\hat{\pi}$. For the sake of simpler notation we relabel $(b_{1}',b_{1}'',...,b_{N}',b_{N}'')$ as $(b_{1},b_{2},...,b_{2N})$. The relation between the $c$ and $b$ Majorana fermion operators is given by Eq.~(\ref{eq:Q}):
\begin{equation}\label{eq:linear_transf}
c_{i}=\sum_{j}Q_{i,j}b_{j},
\end{equation}
where $Q$ is an orthogonal matrix. Let us consider the set $S_{2N}$ of the permutations of $1,...,2N$. Since the $c_i$ anticommute, we can write the product $c_1c_2 ... c_{2N}$ as a sum over all permutations $\sigma \in S_{2N}$ as follows:
\begin{equation}\label{eq:prod_c_permuted}
\prod_{i=1}^{2N}c_{i}=\frac{1}{(2N)!}\sum_{\sigma }\epsilon(\sigma)\prod_{k=1}^{2N}c_{\sigma(k)}, 
\end{equation}
where $\epsilon(\sigma)$ is the sign of permutation $\sigma$. By using Eq.~(\ref{eq:linear_transf}) we can write $\prod_{k=1}^{2N}c_{\sigma(k)}$ as
\begin{eqnarray}
&& \sum_{i_{1},...,i_{2N}} Q_{\sigma(1),i_{\sigma(1)}}...Q_{\sigma(2N),i_{\sigma(2N)}} \, b_{i_{\sigma(1)}}...b_{i_{\sigma(2N)}} \nonumber\\
 = && \sum_{i_{1},...,i_{2N}} Q_{1 , i_1}... Q_{2N , i_{2N}} \,  b_{i_{\sigma(1)}}... b_{i_{\sigma(2N)}},
\end{eqnarray}
where the numerical factor $\prod_k Q_{k , i_k}$ only depends on the values of $i_k$, and not on the permutation $\sigma$. This allows to express Eq.~(\ref{eq:prod_c_permuted}) in the following form
\begin{equation}\label{eq:wef1}
\frac{1}{(2N)!}\sum_{i_{1},..,i_{2N}} \prod_{k=1}^{2N}Q_{k , i_{k}}\left( \sum_{\sigma} \epsilon(\sigma) b_{i_{\sigma(1)}} ... b_{i_{\sigma(2N)}} \right).
\end{equation}
It is not difficult to check that the sum over $\sigma$ gives zero if the values of two of the indexes $i_k$ are equal. When the $i_k$ are all distinct, they are a permutation of $1,...,2N$: $i_k=\sigma'(k)$ and $i_{\sigma(k)}=\sigma'(\sigma(k))$. \color{black} Furthermore, we can anticommute the Majorana operators to the canonical order $b_1 b_2 ... b_{2N}$, which introduces the sign $\epsilon(\sigma)\epsilon(\sigma')$.\color{black} This leads to
\begin{equation}\label{eq:wef2}
\frac{1}{(2N)!}\left( \sum_{\sigma' \sigma} \epsilon(\sigma') \prod_{k=1}^{2N}Q_{k , \sigma'(k)}  \right)  b_{1}b_2 ... b_{2N}.
\end{equation} 
where the $(2N)!$ is canceled by the sum over $\sigma$. The remaining factor is simply the determinant of $Q$, and the following relation is obtained:
\begin{equation}
\prod_{i=1}^{2N}c_{i}=\det(Q)\prod_{i=1}^{2N}b_{i},
\end{equation}
from which Eq.~(\ref{c_product}) directly follows.

\section{Derivation of Eq.~(\ref{eq:P0})\label{app2}}
In this appendix we present details of the derivation of Eq.~(\ref{eq:P0}). From Eq.~(\ref{P_factorization}) we know that
\begin{eqnarray}
2\mathcal{P}_{0}=1+\prod_{i=1}^{2N}D_{i},
\end{eqnarray}
with gauge operators $D_{i}=b_{i}^{x}b_{i}^{y}b_{i}^{z}c_{i}$. The product of all gauge operators can then be rewritten in terms of Majorana operators as
\begin{equation}
\prod_{i=1}^{2N}D_{i}=b_{1}^{x}b_{1}^{y}b_{1}^{z}c_{1} \, \ldots  \, b_{2N}^{x}b_{2N}^{y}b_{2N}^{z}c_{2N}.
\end{equation}
where $N=L_1 L_2$. In order to pair the $b$ operators and form the corresponding $u$ operators, we first move all the $c$ operators to the right of the $b$ operators. The fermionic anticommutation relations then lead to a phase factor of $(-1)^{\phi_{1}}$ with $\phi_{1}=3N(2N-1)$:
\begin{equation}
\prod_{i=1}^{2N}D_{i}=(-1)^{\phi_{1}}b_{1}^{x}b_{1}^{y}b_{1}^{z} \, \ldots  \, b_{2N}^{x}b_{2N}^{y}b_{2N}^{z}  \prod_{l=1}^{2N} c_{l} .
\end{equation}
Since $b_{i}^{x}$ and $b_{i+1}^{x}$ are always separated by two fermionic operators, we can move all the $b^{x}$ to the left without introducing any phase factor. We then group together all the $b^{y}$ operators to the right of the $b^{x}$, at the cost of an additional phase $(-1)^{\phi_{2}}$ where $\phi_{2}=N(2N-1)$ [notice that $(-1)^{\phi_1+\phi_2}=1$]: 
\begin{equation}
\prod_{i=1}^{2N}D_{i}=\prod_{i=1}^{2N}b_{i}^{x}\prod_{j=1}^{2N}b_{j}^{y}\prod_{k=1}^{2N}b_{k}^{z}\prod_{l=1}^{2N}c_{l},
\end{equation}

For the sake of clarity we explicitly write in this appendix each $u$ operator as $u^{\alpha}$ where $\alpha$ refers to the link associated to $u$: $\hat{u}_{ij}=\hat{u}_{ij}^{\alpha_{ij}}=ib_{i}^{\alpha_{ij}}b_{j}^{\alpha_{ij}}$. \color{black} In order to pair the $b^{\alpha}$ ($\alpha=x,y,z$) operators into $u^{\alpha}$, it is necessary to fix the correspondence between $i=1,...,2N$ and the sites of the lattice. We choose here the labeling defined by Eq.~(\ref{eq:labeling}) and illustrated in Fig.~\ref{fig:HoneycombTorus} for a particular case ($L_{1}=4$ and $L_{2}=M=2$). It is then straightforward to see that
\begin{equation}
\prod_{k=1}^{2N}b_{k}^{z}=\frac{(-1)^{\phi_{3}}}{i^N}\prod_{\langle m,n\rangle}u_{mn}^{z}
\end{equation}
where the factor $1/i^{N}$ is from the definition of $u_{ij}^{\alpha}=i b_{i}^{\alpha}b_{j}^{\alpha}$ and $\phi_{3}=N$ arises from the convention of specifying $u_{ij}$ with $i\in A$. Therefore, the $u_{mn}^z$ above have always the form $u^z_{i+1 \, i}$. 

It is also not difficult to rearrange the $b^x$ operators: 
\begin{equation}
\prod_{i=1}^{2N}b_{i}^{x}=\frac{(-1)^{\phi_{4}}}{i^N}\prod_{\langle m,n\rangle}u_{mn}^{x}
\end{equation}
where $\phi_{4}=L_{2}$. Note that the phase $\phi_{4}$ arises because of the boundary conditions along ${\bf e}_{1}$. To form the $u_{1 \, 2L_1}^x$ operator ($u^x_{18}$ in Fig.~\ref{fig:HoneycombTorus}) one has to move $b_1^x$ after $b_{2L_1}^x$, which introduces a $(-1)$ factor. This procedure has to be repeated $L_2$ times, for all pairs of the form $b^x_{2 n L_1}b^x_{2L_1(n-1)+1}$ with $n=1,2,\ldots, L_2$ (in Fig.~\ref{fig:HoneycombTorus}, these are $b^x_{8}b^x_{1}$ and $b^x_{16}b^x_{9}$). Hence, the factor $(-1)^{L_2}$ arises. 

Finally, in order to identify $\prod_{j=1}^{2N}b_{j}^{y}$ as a product of $u^{y}$, we first decompose it in $L_{2}$ products of $2L_{1}$ terms:
\begin{equation}\label{eq:by}
\prod_{j=1}^{2N}b_{j}^{y}=\prod_{n=1}^{L_{2}}\mathcal{T}_{n},
\end{equation} 
where $\mathcal{T}_{n}=\prod_{j=1}^{2L_1}b^y_{2L_1(n-1)+j}$ (e.g., $\mathcal{T}_{1}=b_1^y b_2^y \ldots b_8^y$ for Fig.~\ref{fig:HoneycombTorus}). As a first step, we rewrite each $\mathcal{T}_{n}$ by moving all the $b_{i}^{y}$ with odd $i$ on the left side (and keeping them in increasing order of $i$) while rearranging the $b_{i}^{y}$ with even $i$ (now on the right side of each $\mathcal{T}_{n}$) in decreasing order. These operations do not introduce any additional phase factor in the final expressions. For example, in the case of Fig.~\ref{fig:HoneycombTorus} we can write $\mathcal{T}_{1}=b_1^y b_3^y b_5^y b_7^y b_8^y b_6^y b_4^yb_2^y$. The advantage of this ordering is that most pairs of $b^y$ operators are now straightforward to form. For example, the pairs $b^y_{2}b^y_{9}$, $b^y_{4}b^y_{11}$, $\ldots $ of Fig.~\ref{fig:HoneycombTorus} are now easily formed from the rearranged string (\ref{eq:by}), of the form $\ldots  b_6^y b_4^yb_2^y b_9^y b_{11}^y b_{13}^y \ldots $. The only difficulty is a remaining product of $b^y_i$
\begin{equation}\label{eq:remaining_by_string}
(b_1^y b_3^y \ldots b^y_{2L_1-1})(b^y_{2N}b^y_{2N-2} \ldots b^y_{2N-2(L_1-1)}),
\end{equation}
which requires some care in pairing to account of the toric boundary conditions. The rearrangement of Eq.~(\ref{eq:remaining_by_string}) introduces the phase factor $(-1)^{\phi_5}$ in the final expression:
\begin{equation}
\prod_{j=1}^{2N}b_{j}^{y}=\frac{(-1)^{\phi_{5}}}{i^N}\prod_{\langle m,n\rangle}\hat{u}_{mn}^{y},
\end{equation}
where $\phi_{5}=M(L_{1}-M)+L_{1}$.

Collecting all the terms, we finally obtain
\begin{equation}
\prod_{i=1}^{2N}D_{i}=\frac{(-1)^{\phi_3+\phi_4+\phi_5}}{i^{3N}}\prod_{\langle i,j\rangle}\hat{u}_{ij}\prod_{l=1}^{2N}c_{l}.
\end{equation}
With the aid of Eq.~(\ref{c_product}) we then find
\begin{equation}
2\mathcal{P}_{0}=1+(-1)^{L_{1}+L_{2}+M(L_{1}-M)}\det(Q^{u})\hat{\pi}\prod_{\langle i,j\rangle}\hat{u}_{ij},
\end{equation}
which is Eq.~(\ref{eq:P0}) of the main text.


\begin{thebibliography}{100}

\bibitem{KitaevHoney}
A. Kitaev, Ann. Phys. \textbf{321}, 2 (2006).

\bibitem{Nussinov2008}
H.-D. Chen and Z. Nussinov, J. Phys. A: Math. Theor. \textbf{41}, 075001 (2008).

\bibitem{Vidal2008} 
J. Vidal, K. P. Schmidt, and S. Dusuel, Phys. Rev. B {\bf 78}, 245121 (2008).

\bibitem{KellsPRB2009}
G. Kells, J. K. Slingerland, and J. Vala, Phys. Rev. B \textbf{80}, 125415 (2009).

\bibitem{KitaevMott}
G. Jackeli and G. Khaliullin, Phys. Rev. Lett. \textbf{102}, 017205 (2009).

\bibitem{KitaevPRL2011}
K. S. Tikhonov, M. V. Feigel'man, and A. Yu. Kitaev, Phys. Rev. Lett. \textbf{106}, 067203 (2011).

\bibitem{KitaevToric}
A. Kitaev, Ann. Phys. \textbf{303}, 2 (2003).

\bibitem{ChesiPRA}
S. Chesi, B. R\"othlisberger, and D. Loss,  Phys. Rev. A {\bf 82}, 022305 (2010).

\bibitem{MemoryPRB}
F. L. Pedrocchi, S. Chesi, and D. Loss, Phys. Rev. B \textbf{83}, 115415 (2011).

\bibitem{OpticalLattices}
L.-M. Duan, E. Demler, and M. D. Lukin, Phys. Rev. Lett. \textbf{91}, 090402 (2003).

\bibitem{Nori}
J. Q. You, X.-F. Shi, X. Hu, and F. Nori, Phys. Rev. B \textbf{81}, 014505 (2010). 

\bibitem{YaoPRL2007}
H. Yao and S. A. Kivelson, Phys. Rev. Lett. \textbf{99}, 247203 (2007).

\bibitem{YaoPRL2009}
H. Yao, S.-C. Zhang, and S. A. Kivelson, Phys. Rev. Lett. \textbf{102}, 217202 (2009).

\bibitem{YaoPRL2010}
H. Yao and X.-L. Qi, Phys. Rev. Lett. \textbf{105}, 080501 (2010).

\bibitem{Willans2011}
A. J. Willans, J. T. Chalker, and R. Moessner, Phys. Rev. B \textbf{84}, 115146 (2011).

\bibitem{Pachos2008}
V. Lahtinen, G. Kells, A. Carollo, T. Stitt, J. Vala, and J. K. Pachos, Ann. Phys. \textbf{323}, 2286 (2008).

\bibitem{FiniteSizeEffect2009}
G. Kells, N. Moran, and J. Vala, J. Stat. Mech.: Theory Exp. (2009) P03006.

\bibitem{ChaloupkaPRL2010}
J. Chaloupka, G. Jackeli, and G. Khaliullin, Phys. Rev. Lett. \textbf{105}, 027204 (2010).

\bibitem{Trebst2011}
H.-C. Jiang, Z.-C. Gu, X.-L. Qi, and S. Trebst, arxiv:1101.1145 (2011).

\bibitem{ShankarPRL2007}
G. Baskaran, S. Mandal, and R. Shankar, Phys. Rev. Lett \textbf{98}, 247201 (2007).

\bibitem{PachosPRL2008}
G. Kells, A. T. Bolukbasi, V. Lahtinen, J. K. Slingerland, J. K. Pachos, and J. Vala, Phys. Rev. Lett. \textbf{101}, 240404 (2008).

\bibitem{MandalPRB2009}
S. Mandal and N. Surendran, Phys. Rev. B \textbf{79}, 024426 (2009).

\bibitem{Schmidt2010}
M. Kamfor, S. Dusuel, J. Vidal, and K. P. Schmidt, J. Stat. Mech.: Theory Exp. (2010) P08010.

\bibitem{Taylor2011}
H. Xu and J. M. Taylor, arXiv:1104.0024 (2011).

\bibitem{Lahtinen2001New}
V. Lahtinen, New J. Phys. \textbf{13}, 075009 (2011).

\bibitem{Kugel}
K. I. Kugel' and D. I. Khomskii, Sov. Phys.-Usp. \textbf{25}, 231 (1982).

\bibitem{Lieb}
E. H. Lieb, Phys. Rev. Lett. \textbf{73}, 2158 (1994).

\bibitem{Macris1996}
N. Macris and B. Nachtergaele, J. Stat. Phys. {\bf 85}, 745 (1996).

\end{thebibliography}
\end{document}